\documentclass{easmod}
\usepackage{graphicx}
\def\Swift{{\textit{Swift}}}

\begin{document}

\title{Keck Observations of 160 Gamma-Ray Burst Host Galaxies} 
\runningtitle{Keck Observations of GRB Host Galaxies} 
\author{Daniel A. Perley}\address{Department of Astronomy, California Institute of Technology,
MC 249-17, \\
1200 East California Boulevard,
Pasadena CA 91125, USA}
\author{Joshua S. Bloom}\address{Department of Astronomy, University of California,
  Berkeley, CA 94720, USA}
\author{Jason X. Prochaska}\address{Department of Astronomy and Astrophysics, UCO/Lick Observatory, \\
University of California, Santa Cruz, CA 95064, USA}
\begin{abstract}
We present a preliminary data release from our multi-year campaign at Keck Observatory to study the host galaxies of a large sample of Swift-era gamma-ray bursts via multi-color ground-based optical imaging and spectroscopy.  With over 160 targets observed to date (and almost 100 host detections, most of which have not previously been reported in the literature) our effort represents the broadest GRB host survey to date.  While targeting was heterogeneous, our observations span the known diversity of GRBs including short bursts, long bursts, spectrally soft GRBs (XRFs), ultra-energetic GRBs, X-ray faint GRBs, dark GRBs, SN-GRBs, and other sub-classes.  We also present a preview of our database (currently available online via a convenient web interface) including a catalog of multi-color photometry, redshifts and line ID's.  Final photometry and reduced imaging and spectra will be available in the near future.
\end{abstract}
\maketitle
\section{Introduction}
Studies of the host galaxies of cosmic gamma-ray bursts have been slow to catch up with the revolution in the field sparked by the 2004 launch of the \Swift\ satellite (Gehrels et al.\ 2004).  While the large numbers of GRBs detected by \Swift\ have enabled rapid strides in the understanding of the early behavior and multiwavelength evolution of GRB afterglows (as well as setting records for the brightest and most distant such events; see Gehrels et al. 2009 for a review of \Swift\ GRB results), host-galaxy follow-up remains a quite observationally-intensive endeavor, accessible only to large ground-based telescopes or major space observatories.   The typical \Swift\ long-duration GRB is at a redshift of $z>2$ (Jakobsson et al. 2006); the typical host is $R=25$ mag and often fainter (Hjorth et al. 2012).   So while most pre-\Swift\ GRBs with afterglow localizations also have known host galaxies, the number of published hosts in the \Swift\ era remains quite limited in comparison to the number of GRBs that have occurred since the \Swift\ launch (over 700).  Host spectroscopy is even more challenging to acquire.

Nevertheless, host galaxy observations provide a wealth of information that cannot be gleaned by other means---the integrated properties of the galaxy (mass, luminosity, age, physical size, and so on) are essential to a proper understanding of the gamma-ray burst progenitor and its cosmological context.  In fact, for a significant fraction of bursts, host galaxy observations provide the \emph{only} way to understand the burst environment in any detail or to measure redshifts.   In particular, an absorption redshift has never been derived from a short burst afterglow, and ``dark'' gamma-ray bursts lack (by definition) a bright afterglow.  Indeed, about 75\% of all Swift GRBs have no afterglow redshift.

Starting in 2005 (shortly after the launch of \Swift), we have been continuously conducting deep observations of gamma-ray burst positions to produce a legacy sample of gamma-ray burst host galaxies that is both \emph{large} enough to expand on pre-\Swift\ results in a meaningful way and \emph{diverse} enough to incorporate not just ``ordinary'' bright long-duration bursts but also to enable the detailed study of interesting GRB subclasses that were hardly constrained by pre-\Swift\ studies at all.  In this summary, we present a brief outline of our host discovery program and a preview of early science results.

\section{Program Summary and Observations}

Our observations do not constitute a single homogeneously-defined survey, but rather represent a combination of smaller projects.  Most observations were conducted between 2005--2010 under a series of proposals (PI J.~Bloom) focusing on \emph{host discovery} and basic characterization (via the observed-frame optical color), and placing \emph{redshift constraints}, in particular to rule out a large high-$z$ fraction that was suggested in some early works (e.g., Bromm \& Loeb 2002).  These observations are supplemented by observations from a number of other researchers (PIs Kulkarni, Ofek, Prochaska) on individual observations of interest plus as a few target-of-opportunity observations which were not afterglow-dominated.  Observations continue today, mostly focused on supplementing multi-color photometry and determining photometric redshifts.  Multiple instruments were employed but the large majority of observations were conducted with the Low Resolution Imaging Spectrograph (LRIS; Oke et al. 1995). 

Nights were scheduled classically, and therefore were subject to a variety of observing conditions (seeing, transmission, lunation, etc.)  For observations on nonphotometric nights on imaging fields without Sloan Digital Sky Suvery calibration data, we separately observed with the 1-meter Nickel Telescope at Lick Observatory and the 60-inch Telescope at Palomar Observatory to obtain calibrations of these fields.

As of December 2012, we have imaged a total of 159 unique GRB fields (excluding observations during heavy clouds, fields with severe contamination at the host position by nearby stars, or observations shortly after the GRB which were afterglow-dominated).  Host galaxies or likely host candidates have been detected in 105 of these cases (Figure 1).   We have acquired spectroscopy (typically relatively shallow integrations of 30--90 minutes per target) for 48 targets leading to 21 redshift measurements, 14 of which were new at the time of observation.

Nearly all hosts were observed in at least two optical filters (usually $g$ plus either $R$ or $I$), although usually not in the NIR.  This means that while we are sensitive exclusively to the young stellar population in all but the closest ($z < 1.0$) hosts and therefore cannot usefully constrain the ages or stellar masses of our sample, we can  constrain the average dust reddening of using the empirical UV-slope method (e.g., Meurer et al. 1999). 

\section{Preview of Results}

While GRB host galaxies are canonically thought of as very blue and nearly dust-free (e.g., Le Floc'h et al 2002),
most of the hosts we detect show evidence for significant reddening: $A_V = 1.0-2.5$ mag is typical for the sample.  A large fraction of the reddest hosts are ``dark'' bursts (Figure 2), consistent with the interpretation of these events as dust-extinguished, but even optically bright bursts often have fairly red slopes.    Of course, this measurement is naturally biased to the most luminous hosts in the sample, which are expected to have higher mean dust attenuations than the more ``canonical'' ultra-faint hosts (e.g.\ the host of GRBs 030329; Gorosabel et al. 2005), which we cannot detect at $z \sim 2$.  Nevertheless, it is clear that very (bolometrically) luminous hosts are relatively common at $z \sim 2$.  

Many short GRBs we have observed show no evidence for a host within the XRT error circle at all, which is curious given that the known short GRB redshift distribution is heavily concentrated at $z<1$ and includes several quite luminous galaxies (Prochaska et al. 2006).  This population of apparently  ``hostless'' events suggests a progenitor that has been ejected far from its original host in some cases (see also Berger 2010).

The hardness of the prompt emission does not appear to correlate in any significant way with the properties of its host.  In particular, spectrally-soft \Swift\ X-ray flashes ($E_{\rm peak}$) generally have blue, star-forming hosts similar to those of harder long-duration GRBs.  

\section{Data Access}

We have placed online at \texttt{http://www.astro.caltech.edu/grbhosts/} an index containing imaging thumbnails of all hosts observed during the project and (for most detected hosts) photometry from the $R$ and $I$ filters as well as a list of measured redshifts and line identifications.  As we complete final calibration checks in the coming year, we plan to augment this website with photometry on the remaining objects and filters.  Reduced and calibrated images and extracted spectra will all be placed online for community use.  Users interested in data on particular events of interest before then are encouraged to contact us for more information.

\acknowledgements
Many people have contributed to the acqusition of observations in our catalog, including S.~Kulkarni, A.~Filippenko, E.~Ofek, R.~Foley, S.~B.~Cenko, A.~Soderberg, N.~Butler, D.~Kocevski, K.~Alatalo, M.~Modjaz, A.~Miller, J.~Silverman, C.~Thoene, M.~Kasliwal, A.~Morgan, C.~Klein, L.~Pollack, J.~Hennawi, J.~Kirkpatrick, S.~Ellison, S.~Wiktorocicz, M.~van Kerkwijk, K.~Clubb, A.~Horesh, and Y.~Cao.  We also wish to thank the excellent support provided by Keck Observatory, as well as by Lick and Palomar Observatories for our calibration program.
\endacknowledgements

\begin{figure}
\centerline{
\includegraphics[scale=0.97,angle=0]{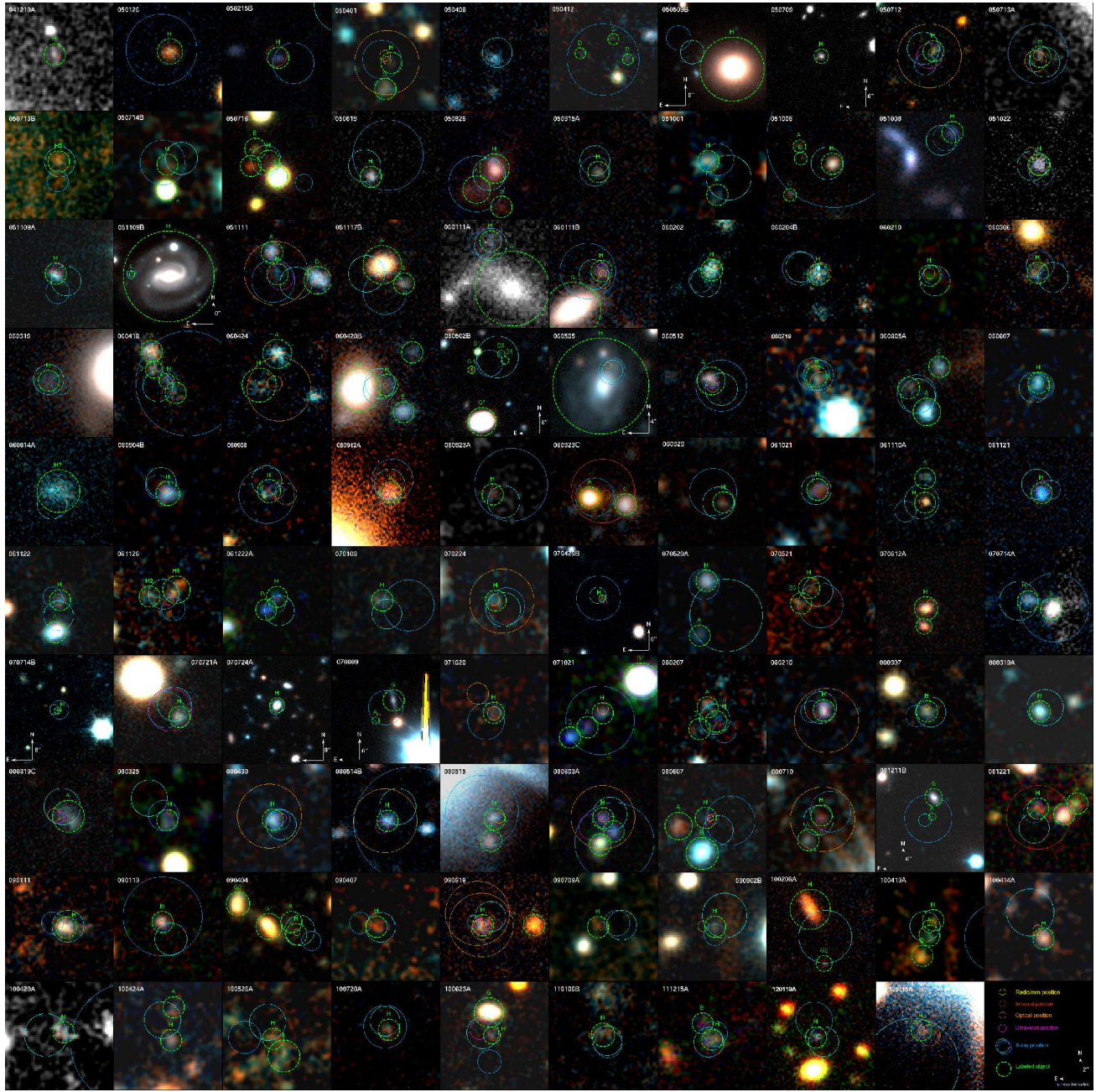}} 
\caption{Mosaic of 99 (out of 105) probable host galaxies detected in the survey.  Host galaxies are identified with an H; afterglow positions and other objects of interest are also marked (see legend at bottom right).  Images are 10$^{\prime\prime}$$\times$10$^{\prime\prime}$ unless labeled otherwise by a scalebar.  A full-resolution version of this figure is online at http://www.astro.caltech.edu/grbhosts/grb2012/detmosaic.png}
\label{fig:detmosaic}
\end{figure}

\begin{figure}
\centerline{
\includegraphics[scale=0.9,angle=0]{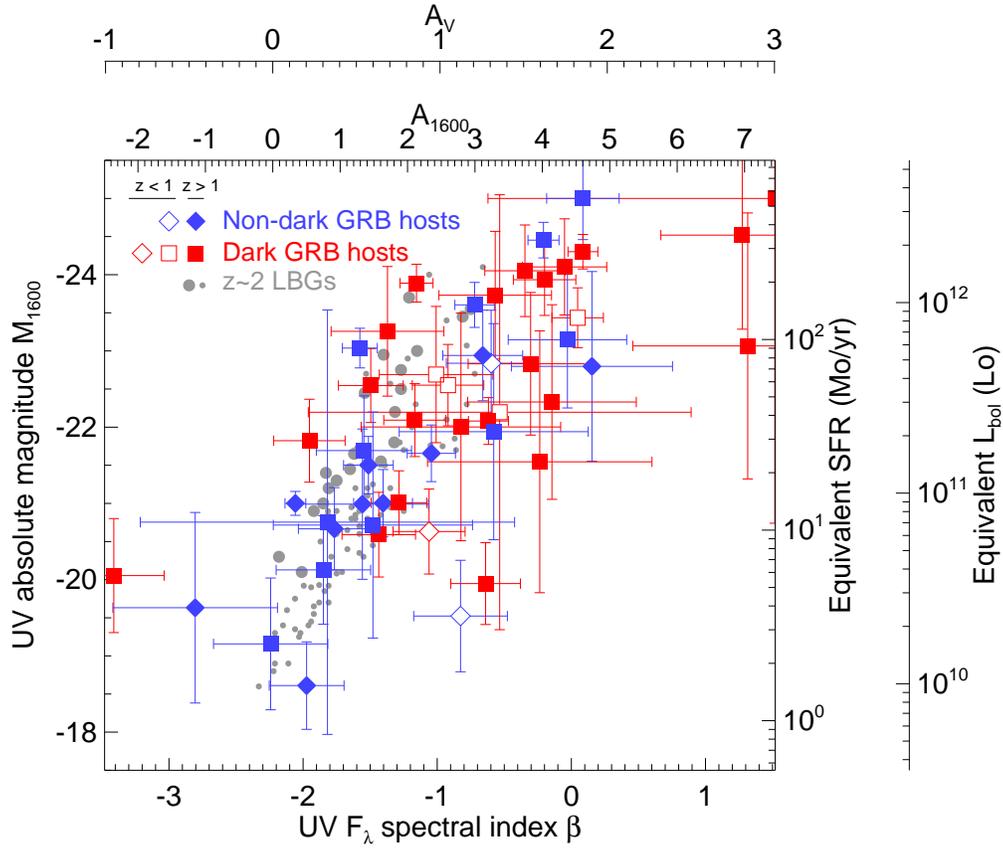}} 
\caption{Dust-corrected UV luminosity versus the UV spectral index $\beta$ (determined from a power-law fit to all filters above the Balmer break) for GRB host galaxies in the sample, plus a sample of pre-Swift hosts with photometry taken from Savaglio et al. 2009 and grbhosts.org.  Among $z>1$ GRBs very luminous host galaxies are actually quite common once dust attenuation is corrected for.  A significant fraction, although not all, of these most luminous hosts correspond to optically-dark or otherwise dust-obscured GRBs.  Also plotted are a sample of field-selected $z\sim2$ galaxies from Meurer et al. 1999.}
\label{fig:mbeta}
\end{figure}



\end{document}